\documentclass{article}

\usepackage{arxiv}

\usepackage[utf8]{inputenc} 
\usepackage[T1]{fontenc}    
\usepackage{hyperref}       
\usepackage{url}            
\usepackage{booktabs}       
\usepackage{amsfonts}       
\usepackage{nicefrac}       
\usepackage{microtype}      
\usepackage{lipsum}
\usepackage{amsmath}
\usepackage{csquotes}
\usepackage{subfigure}
\usepackage{graphicx}

\title{Link Capacity Distributions and Optimal Capacities for Competent Network Performance}

\author{
  Saptarshi Pal \\
  Department of Applied Mathematics\\
  University of Waterloo\\
  N2L3G1, Waterloo, ON, Canada \\
  \texttt{s24pal@uwaterloo.ca} \\
   \And
Ayan Chatterjee \\
  Network Science Institute\\
  Northeastern University\\
  360 Huntington Ave, Boston, MA, USA \\
  \texttt{chatterjee.ay@husky.neu.edu} \\
  \And
Dripto Bakshi \\
  Indian Statistical Institute\\
  Kolkata, West Bengal, India\\
  \texttt{bakshi.dripto@gmail.com} \\
  \And
Amitava Mukherjee \\
  Adamas University\\
  Kolkata, West Bengal, India\\
  \texttt{amukh10@gmail.com} \\
}

\begin{document}
\maketitle

\begin{abstract}
This work addresses the problem of evaluating optimal link capacities of a packet-flow network for the objective of congestion minimization. We present a simple model of packet flow in networks and present a numerical approach to evaluate packet flow probability mass function at any arbitrary edge of the network for a given routing algorithm and traffic rate. We further discuss techniques of assigning optimal capacity at each edge for attaining desired minimized congestion and discuss related trade-offs. Our framework is built around the assumption of Poisson traffic, however the numerical approach fits for any general distribution of packet influx. Lastly, we define metrics of global performance of link capacities allocation and discuss the effect of network structure on capacity allocation and performance.
\end{abstract}

\section{Introduction}
The minimization of the traffic congestion in the communication networks is a well established branch in network theory. The congestion control and its alleviation were achieved through some real time, sophisticated protocols and algorithms like the ones that were  developed  in  the works  by \cite{jacobson1988congestion} and \cite{floyd1993random}. In those works, the protocols and the algorithms were designed solely for computer communication networks. These algorithms incorporated a general approach of taking certain steps to avoid or control  congestion;  the  examples of these steps are i) the window size  reduction in TCP, ii) the exponential back-off strategy in CSMA/CA protocol and iii) the use of fair queuing algorithm in the work of \cite{demers1990analysis}.    
\\In the case of traffic congestion in road networks, the question of congestion control and its minimization had been studied since the discussion about the applicability of the Braess' Paradox problem in road networks became prevalent like in \cite{steinberg1983prevalence}. Other relevant works on road traffic congestion control had been done in other works like \cite{Stefanello,easley2010networks}. These works included how traffic congestion could be eliminated by manipulating and distributing traffic flow using tollbooths (\cite{Stefanello}) and additional edges (\cite{easley2010networks}. A recent work on urban road traffic congestion alleviation using \textit{congestion charging} done in \cite{ye2012research} received also significant traction.
\\In other recent work by \cite{DBLP:journals/corr/abs-1710-00420}, link dimensioning approach was adopted which started by measuring the statistical parameters of the available paths, and the degree of fluctuations in the traffic flow. This was followed by choosing a proper model to fit the traffic volume using lognormal and generalized extreme value distributions. Finally, the optimal capacity for the link could be estimated by deploying the bandwidth provisioning approach. In another work by \cite{Atamtark}, traffic flow in IP networks was modelled with heavy-tailed distributions such as lognormal and GEV. This paper provided an approach of avoiding congestion by increasing link bandwidth by 30\% above the allocated average link bandwidth. Another work by \cite{4804330} had used similar approach with more generalized traffic distributions, and had also defined congestion in networks analytically. \cite{Yang_1} used nonlinear consensus congestion control algorithm  via Lyapunov function, which also included dynamic bandwidth allocation in communication network.
The definition of congestion is not global in nature but is related to the network under consideration. Also, it is easy to note from the literature cited above that the congestion control algorithms are dependent on the mechanism of traffic flows in network. All the works regarding congestion  control  in  various  types of networks introduced reactive strategies which manipulated the system variables e.g. window size in TCP, flow control rate in CSMA/CA, queuing  algorithms in communication networks, and charging congestion,  eliminating  critical edges in  road networks. Unlike  the works that have  been  discussed  in  the  earlier paragraphs, the \textbf{original contribution of this paper is the proposal} of a strategy that states a proactive algorithm to evaluate the optimal link capacities for a traffic network such that congestion can always be kept under a required (design) threshold, without the introduction of any reactive strategies as stated earlier. Indeed,  this  work  will be subject to the nature of the traffic flow that is assumed, and will also be contingent to the contextual definition of congestion. This strategy, that we implement for evaluating and allocating the capacities of edges in the networks (links and edges are interchangeably used from now onward), assures that the possibility of congestion appears minimally in networks where any routing policies are employed. This work discusses the behaviour of packetized traffic in the communication networks, for example, and reiterates the definition of the congestion given as follows: a link in the network is said to be congested during a  time  interval  only  if  the  number  of  packets  flowing  through the link is greater than its assigned capacity. For the sake of simplicity, we consider communication networks, as an example, to explain our strategy. Our proposed generic strategy will be applicable for any networks like road networks etc.
\\The backbone of our strategy lies in evaluating the probability mass function of  the  number  of packets flowing  through  any  link  in  the network  within  a time interval;it is  given the  necessary underlying information about the transport process used in the network.  The necessary information includes the followings:  (a) the input about the probability of the packet transfer between any pair of nodes in the network (b) the statistical distribution of the number of packets sent between a pair of nodes in the network if and only if  they agree to  send  packets between themselves in  the  first  place  (c)  the  use of routing algorithm and (d) the graph topology corresponding to the network for which we are evaluating the capacities of the links.  Apart from these necessary information, we also require an input of the local performance criteria for each link that define the threshold for congestion in a link in the network.  Details about this local performance criteria is explained later in section 4.
\\In this work, the strategy that we develop for finding the optimal link capacity can be implemented for real networks that follow any given routing protocols (algorithms) (e.g. open shortest path first (OSPF) etc.) for the transmission of packets. A real-world example is road networks supporting a system of ride-sharing  cabs that traverse through a geographical area, say a city, which is a network of streets. Applicability of our strategy holds true even under sparse minimal changes in the routes.  In this example, the capacities of links refer to a number of parameters at the same time, some of which are - the number of street lanes, traffic regulation time at the intersections of the streets etc. Another example of an applicable network is a network of postal services.  Link capacity, in this example, refers to the total capacity of the transporters that take the job of transporting packages from one transportation center to the other.  \\
The paper is split into six sections. In the second  section, we present the traffic flow model where we describe in detail the scheme of traffic flow that we abide by throughout the paper.  Moreover, the model, we describe in the second  section, is  later  used  to  design  a  stochastic  simulation  model which,  in turn, is used to assess the performance and accuracy of our algorithm.  In the third section, we define and describe the mathematical background for this work, and hence introduce a computational approach to evaluate, with a certain degree of error, the probability mass function of the number of packets flowing through all links in the network.  In the fourth section, we describe the  nature  of  the  local  performance  criteria  for  design, and  discuss  how  we evaluate the link capacity for each link in the network from the probability mass function  estimated  in  the  previous section  and  the  local  performance criteria set for each link.  In the fifth section, we define global performance metrics that we use to analyze the performance of our strategy from the perspective of the graph as a whole.  We measure the global performance metric using  the  stochastic simulation  model  which is discussed  in  the  second  section.  We observe and comment on the evaluated performance metrics.  We also show the nature of changes in these performance metrics with the changes in graph (network topology) parameters like mean degree, mean centrality, degree standard deviation and centrality standard deviation. In the sixth  section, we discuss possible future works stemming from the line of work presented in this paper.

\section{Traffic Flow Model}
In this section, we define the traffic flow model that we follow throughout this work. Apart from describing the underlying traffic flow model in the network in question, this model also develops the stochastic simulation scheme which we use to computationally evaluate the global performance metric in a later part of the paper. 
\\
Before describing the model in detail, let us begin by defining the network as an undirected connected graph, $G$=$(V,E)$, where $V$ is the set of nodes representing the set of sources and destinations in the network. All the nodes in $V$ have the capability to generate packets, receive packets and route packets towards their destinations. $E$ represents the set of all bidirectional edges of the network. An edge carries packets between the nodes that it connects in the graph. Since a source in the network will choose any other node in the network as a destination for its transmission of packets, a packet sent by a node traverses through more than one nodes before reaching its destination (i.e., the destination node would not be its immediate neighbour). The traffic flow model that we propose has two separate sections namely: \textbf{packet generation model} and \textbf{packet routing model}. 
\\
Before defining these two sections of traffic flow model, let us define a term called \textbf{time frame} that we use frequently while describing these sections. A time frame is defined as a fixed time span of duration $T$ such that:$$T = t_p + t_r$$
where $t_p$ is the maximum processing time of packets at nodes i.e., the time taken to generate the packets for transmission including queuing time and $t_r$ is the maximum transmission time for flowing packets from a source to a destination in the network. In a time window $T$, $t_p$ precedes $t_r$, and it is generally negligible in comparison to the propagation time for most networks, especially in computer networks, where moderate or negligible queuing is observed. All the nodes in the network generate packets according to the \textit{packet generation model} during the time interval $t_p$. The packets generated at their respective source nodes in the network during this interval, traverse through the network towards their destinations according to \textit{packet routing model} within the time interval $t_r$. Since, generally $t_p << t_r$, a whole time window $T$ is approximately the transmission time $t_r$.\\
\textbf{The Packet Generation Model}: The packet generation model is under operation during the first part of the time window $T$, i.e., within the time interval $t_p$. Within this interval, all the steps of the packet generation model occur sequentially as specified below: 
\begin{enumerate}
\item $\forall$ $i,j$ $\in V$ and $i \neq j$, links are created between node $i$ and node $j$ with a fixed predefined probability of $q_{ij}$. Node $i$ is identified as the source and the node $j$ as the destination in the link connected. It is not necessary $q_{ij} = q_{ji}$ or that a link is  between $j$ and $i$ with certainty if link exists between $i$ and $j$.
\item For each source-destination pair $(i,j)$, that has a connected link between them from the previous step, packets are generated for transmission. The number of packets to be sent from node $i$ to node $j$, given the link exists, is drawn from a Poisson distribution with mean $\lambda_{ij}$.
\end{enumerate}
\textbf{The Packet Routing Model}: This model is under operation only during the transmission time interval, which is of duration $t_r$. It defines how the generated packets in the source of a link move across the network towards the destination through that link. Here, either step 1 or step 2 is followed depending upon the condition is satisfied, that are described below. In the case of both the alternatives, the packets reach destination by the end of the time interval $t_r$.
\begin{enumerate}
\item If there exists a single shortest path from a source to a destination within the network then all packets having this pair of source-destination are sent through this shortest path.
\item If multiple shortest paths, say $L$, exist from a source to a destination, any one of these paths is selected randomly with the probability of  $\frac{1}{L}$ . 
\end{enumerate}

\begin{figure}[h!]       
    \includegraphics[width = 0.5 \linewidth]{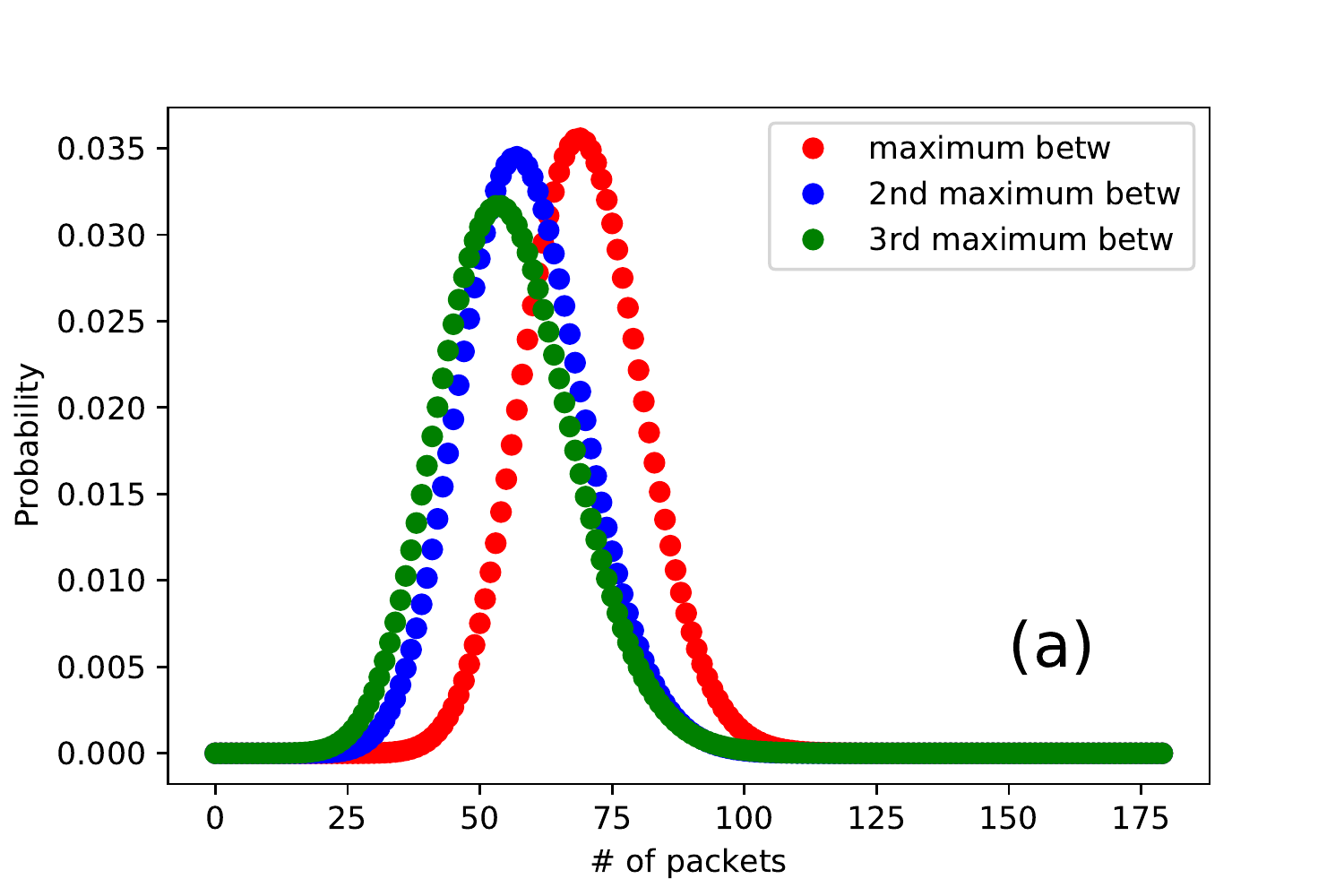} 
    \hspace{30px}
    \includegraphics[width = 0.5 \linewidth]{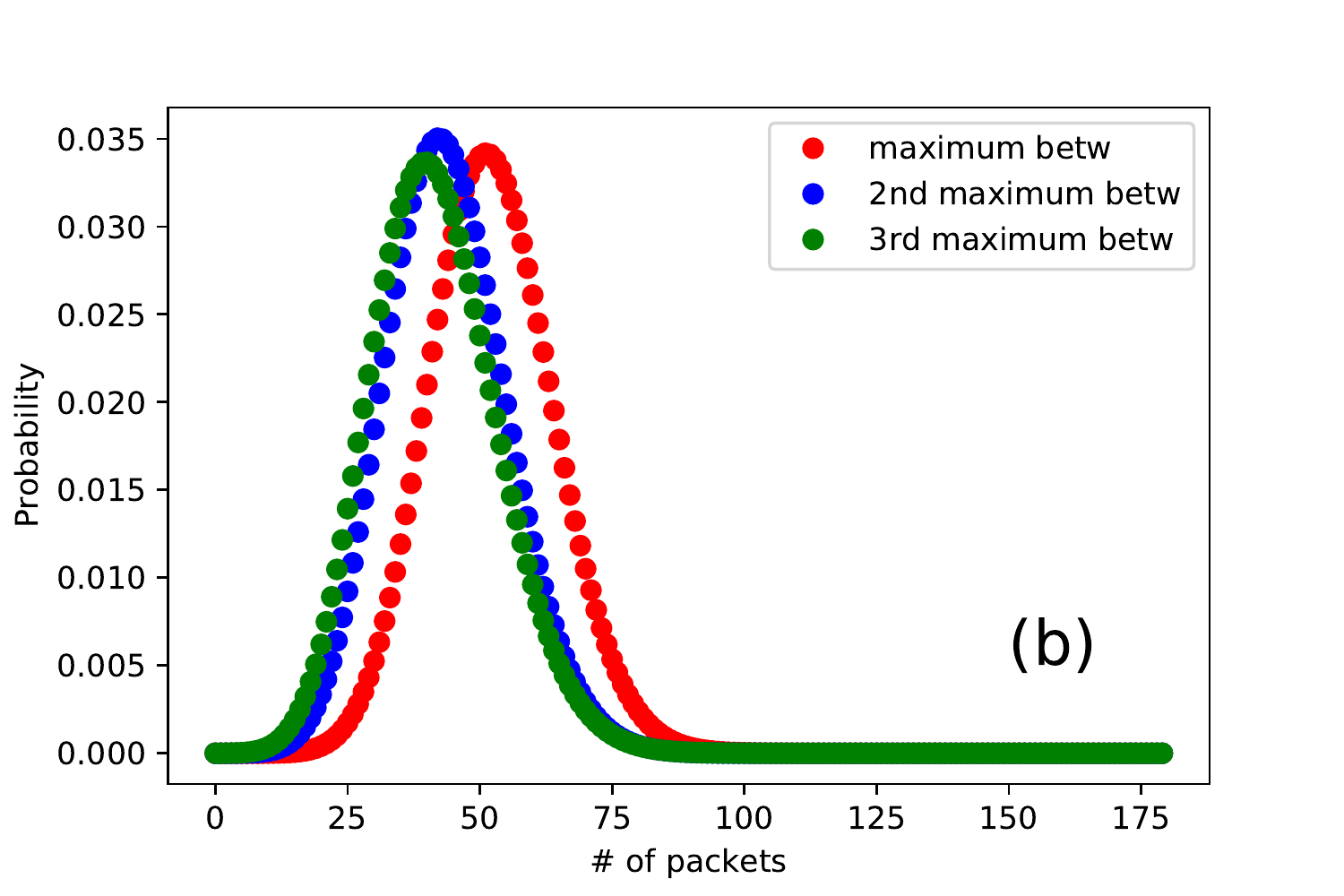}
    \hspace{30px}
    \includegraphics[width = 0.5 \linewidth]{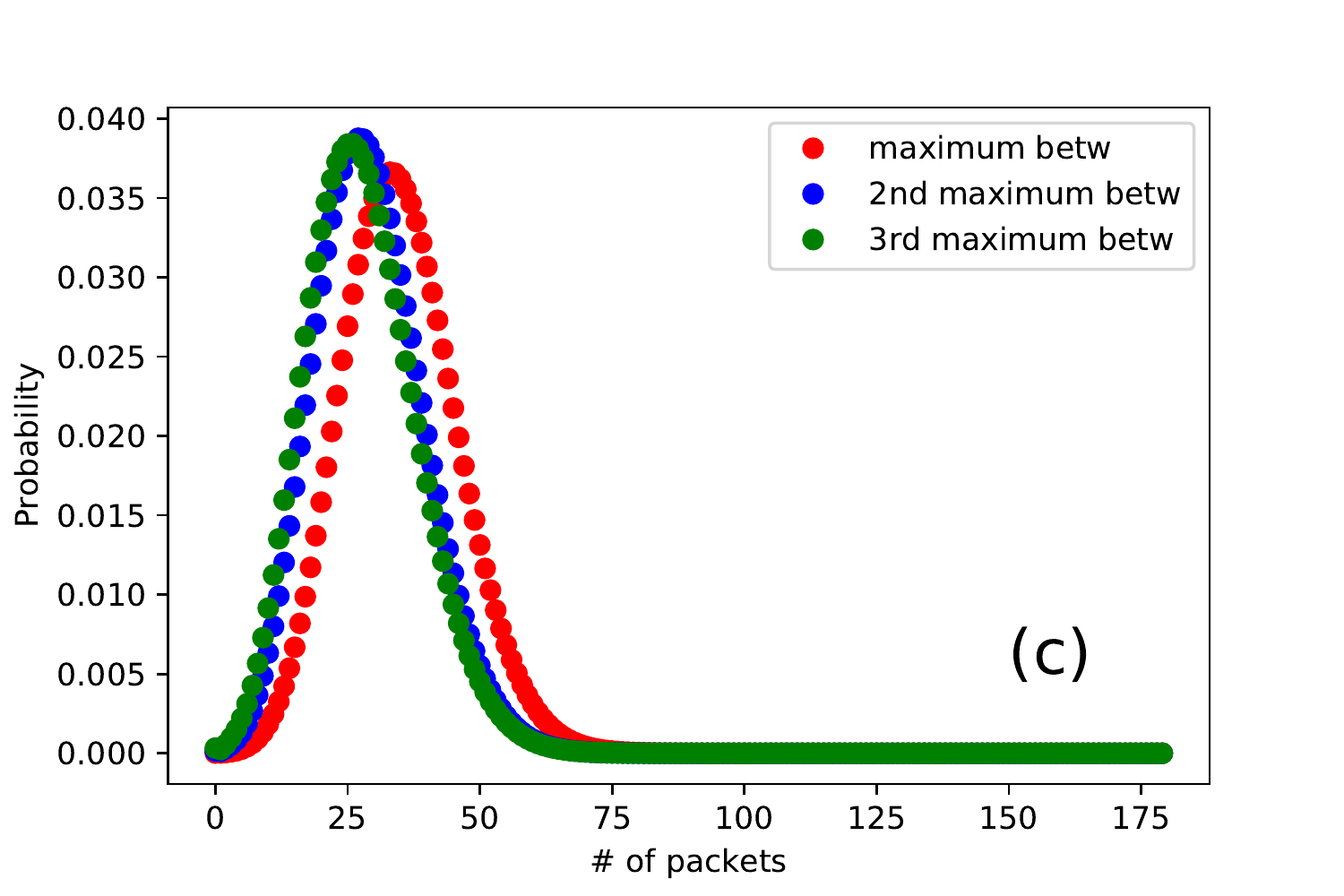}
    \hspace{30px}
    \includegraphics[width = 0.5 \linewidth]{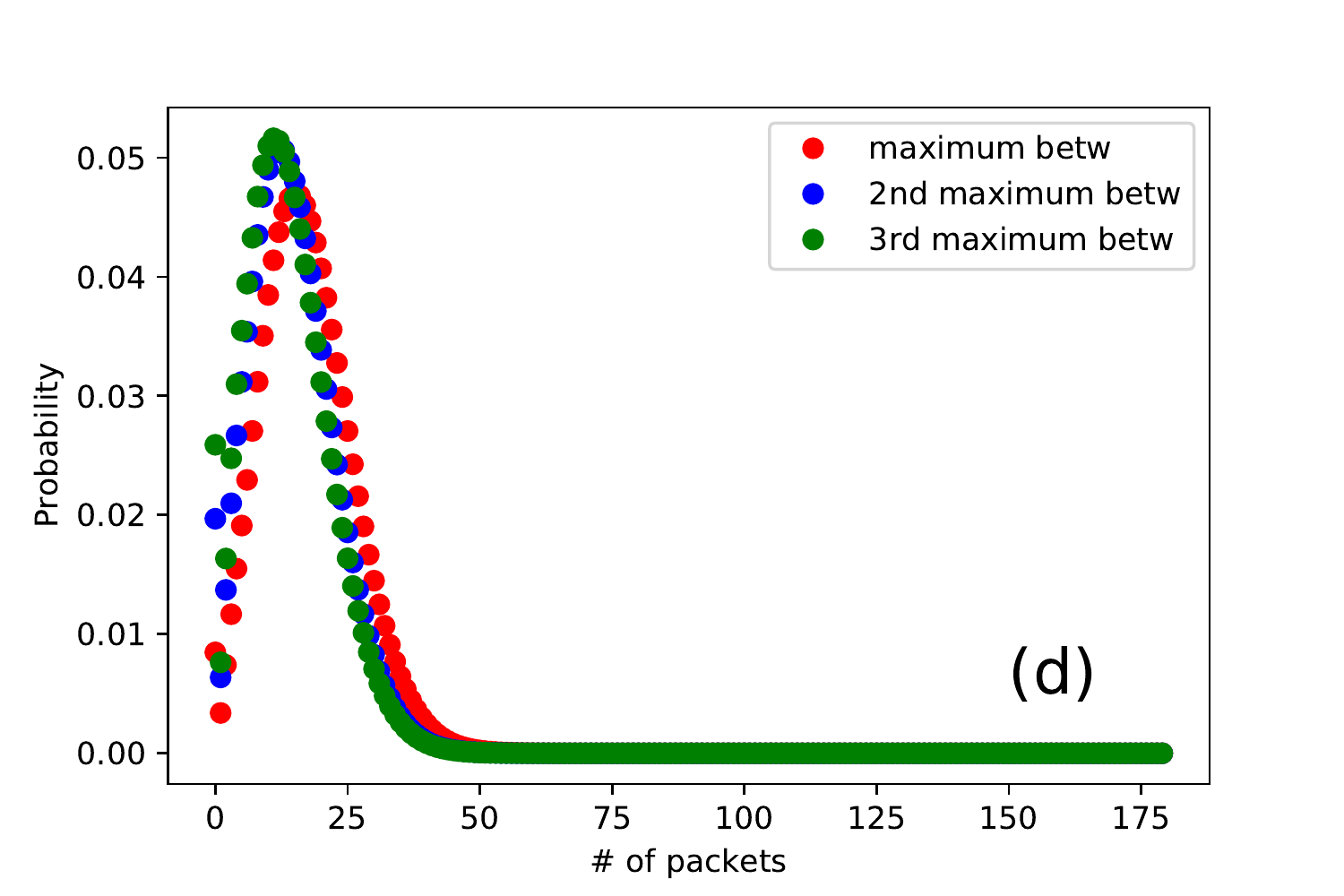}%
    \caption{The evaluated pmf  $\Pi$ is drawn for links having top three highest values of edge betweenness centrality for $q$ = (a) 1, (b) 0.75, (c) 0.5, (d) 0.25. The network is Bar\'{a}basi Albert type with 30 nodes with $\lambda$ = 4. $\Pi(k)$ is plotted from $k$ = 0 to $k$ = 179}
    \label{Fig1}
\end{figure}

Unlike in \cite{PhysRevE.71.026125}, in our traffic model, we do not consider queue formation at any node that lies in the path of transmitting packets. This is because, unlike \cite{PhysRevE.71.026125}, we do not assume that nodes in the network have a finite and limited capacity of processing required for routing. Furthermore, we model in a way where packets traversing towards destination cover more than one hop per time interval. We assume that within the assigned transmission time interval $t_r$, packets generated at their respective sources during the time interval $t_p$, reach their destinations. We assume that the transmission speed is sufficiently large to satisfy the timing condition $t_p << t_r$ in our proposed proactive algorithm. These deviations from the model definition in \cite{PhysRevE.71.026125} are taken into account while determining the optimal link capacities in network.
\section{Evaluation of the Probability Mass Function}
In this section, we discuss in detail how we evaluate the probability mass functions (pmfs) of the number of packets flowing through any link in the network within one time frame based on the traffic flow model discussed in the previous section. In this proposed method, we present a sequence of steps that generate this pmf of packets for any link using a computational approach.
\\Since the successive time frames are totally independent of each other, due to independent processes of packet generations and packet routing being executed between them, the pmf of the number of packets flowing through a link in a given time frame is independent of the time frame for which the pmf is evaluated.  Let us represent this pmf as $\Pi_{ij}$; the subscripts $ij$ indicate that the pmf for the number of packets flowing through the link connecting nodes $i$ and $j$ in the given network topology. 
\\ In order to  describe our proposed method of evaluating the pmfs, we need to define certain terms as follows:
\begin{enumerate}
\item $P(\lambda,k)$= $\frac{\displaystyle \lambda^k e^{-\lambda}}{\displaystyle k!}$, the pmf of the Poisson Distribution.
\item $f_{ij}^{mn}$: The probability that packets sent from node $m$ to node $n$ in the network will pass through the link $ij$.
\item $\Phi^{mn}(\lambda_{mn}, q_{mn}, k)$: The probability that $k$ packets will be sent from node $m$ towards its destination node $n$. It is noted that this probability is a function of $q_{mn}$, which is the probability that a link exists between nodes $m$ and $n$ (as discussed in the previous section).
\item $\Omega_{ij}^{mn}(\lambda_{mn}, q_{mn}, f_{ij}^{mn}, k)$: The probability that $k$ packets, having source node $m$ and destination node $n$,  will pass through the link $ij$.  
\end{enumerate}

Having defined these probabilities, we  develop relation between them which we will require during the computation of the pmf $\Pi_{ij}$. We  express $\Phi^{mn}$ and $\Omega_{ij}^{mn}$ in terms of $\lambda_{mn}$ and $q_{mn}$, for all $m$ and $n$ $\in$ $V$. Subsequently, we derive the equations (1) and (2).
\begin{equation}
\Phi^{mn}(\lambda_{mn}, q_{mn}, k) = 
\begin{cases}
(1-P(\lambda_{mn},0))(1-q_{mn}) + P(\lambda_{mn},0) & \text{if } k = 0 \\
q_{mn}P(\lambda_{mn},k) & \text{otherwise}
\end{cases}
\end{equation} 
Using equation (1) we can derive the expression of $\Omega_{ij}^{mn}$ as: 
\begin{equation}
\Omega_{ij}^{mn}(\lambda_{mn}, q_{mn}, f_{ij}^{mn}, k) = 
\begin{cases}
f_{ij}^{mn}\Phi(\lambda_{mn}, q_{mn}, 0) + 1 - f_{ij}^{mn}& \text{if } k = 0 \\
f_{ij}^{mn}\Phi(\lambda_{mn}, q_{mn}, k) & \text{otherwise}
\end{cases}
\end{equation} 
In equations (1) and (2), the value of k is any non-negative integer.
Now comes the matter of evaluating  $f_{ij}^{mn}$ for any routing algorithm in a given network. As per packet routing model, that is discussed in the previous section, we define $f_{ij}^{mn}$ to be (i) $1$, if there exists only one shortest path from node $m$ to node $n$ and the link $ij$ lies in it, (ii) $0$, if there is only one or more shortest paths from $m$ to $n$ and $ij$ does not lie in any of them, and (iii) $\frac{L_{ij}}{L}$, if $L$ shortest path exists from $m$ to $n$ and $ij$ lies in $L_{ij}$ of them.\\

\begin{figure}[h!]       
    \includegraphics[width = 0.5 \linewidth]{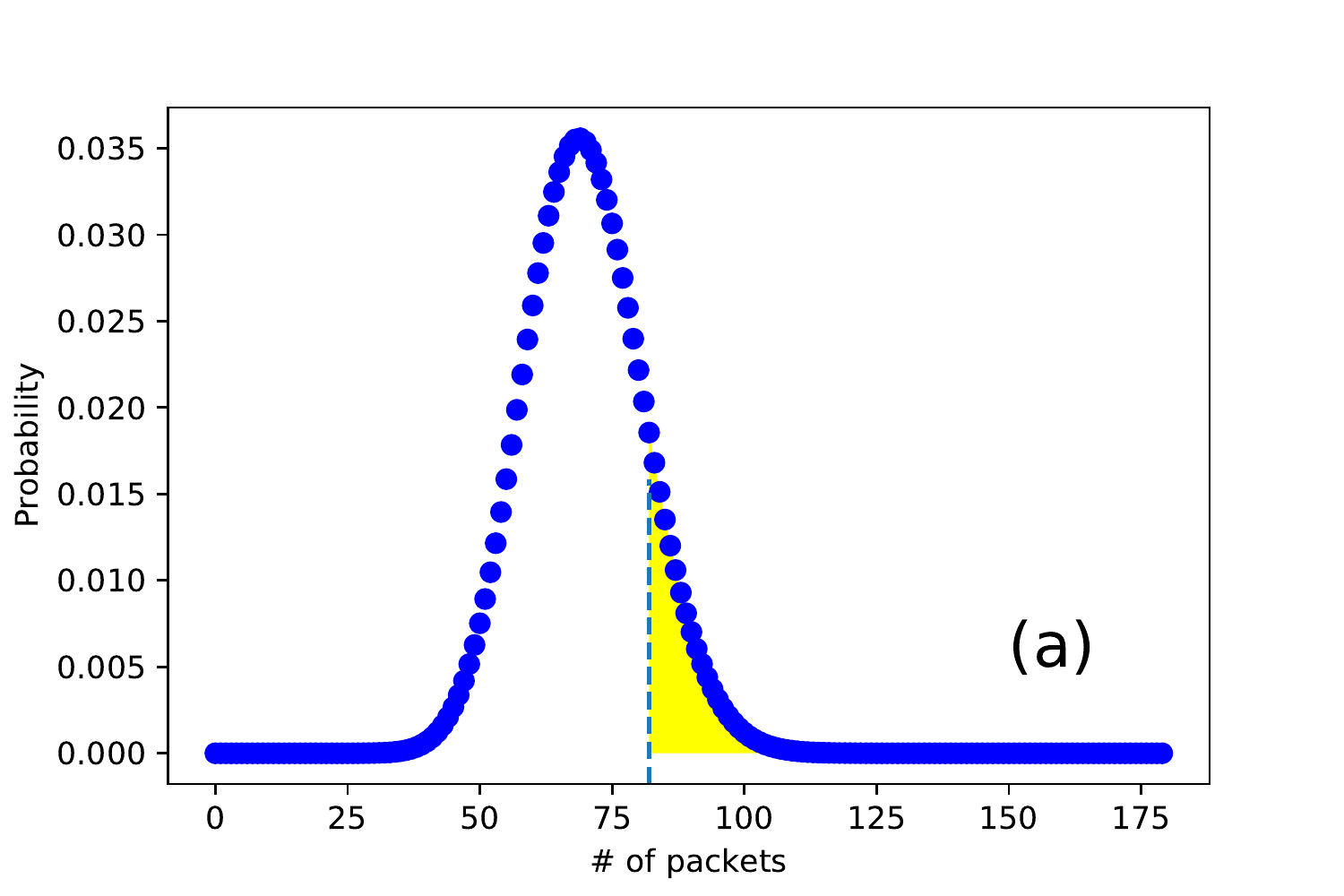} 
    \hspace{30px}
    \includegraphics[width = 0.5 \linewidth]{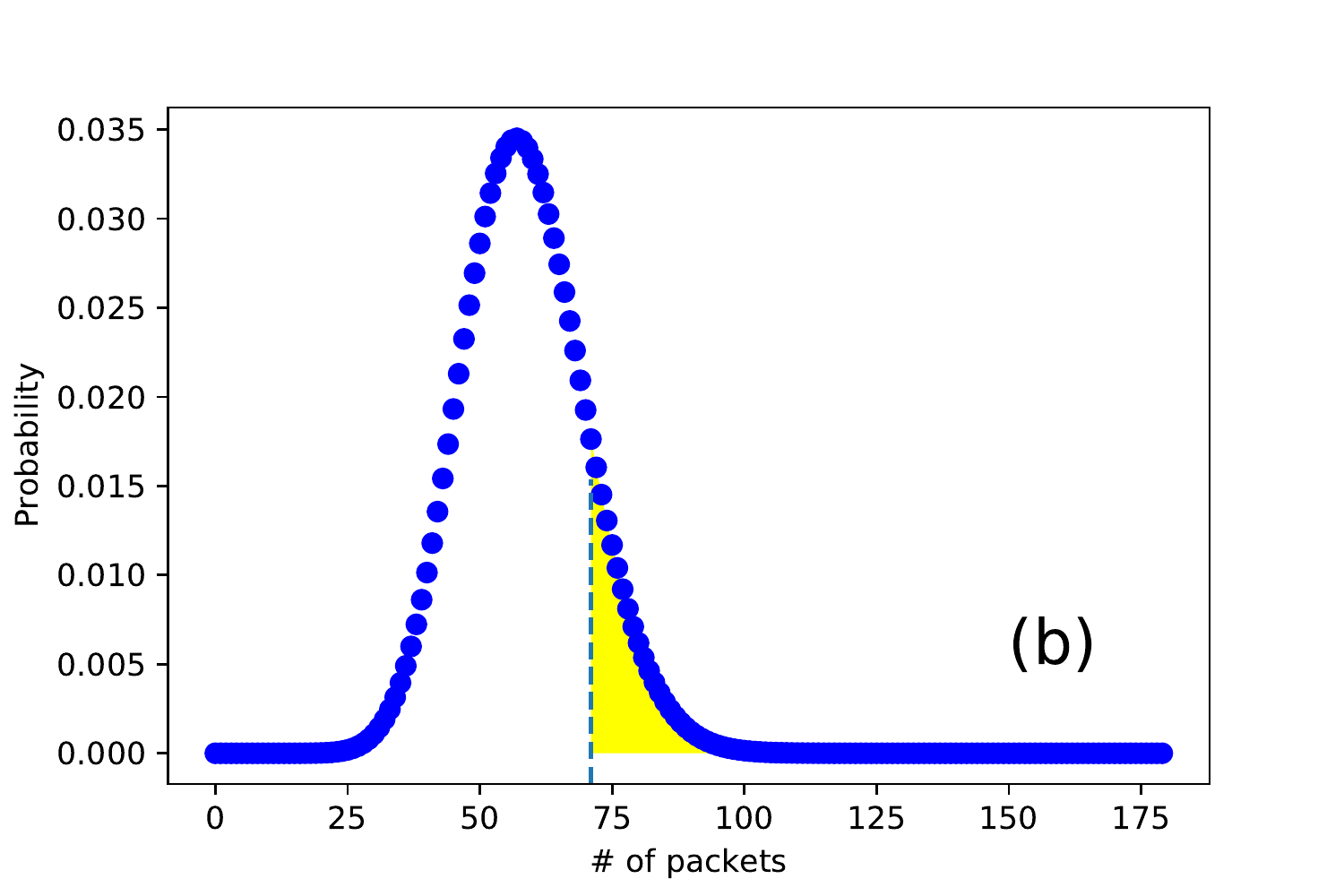}
    \hspace{30px}
    \includegraphics[width = 0.5 \linewidth]{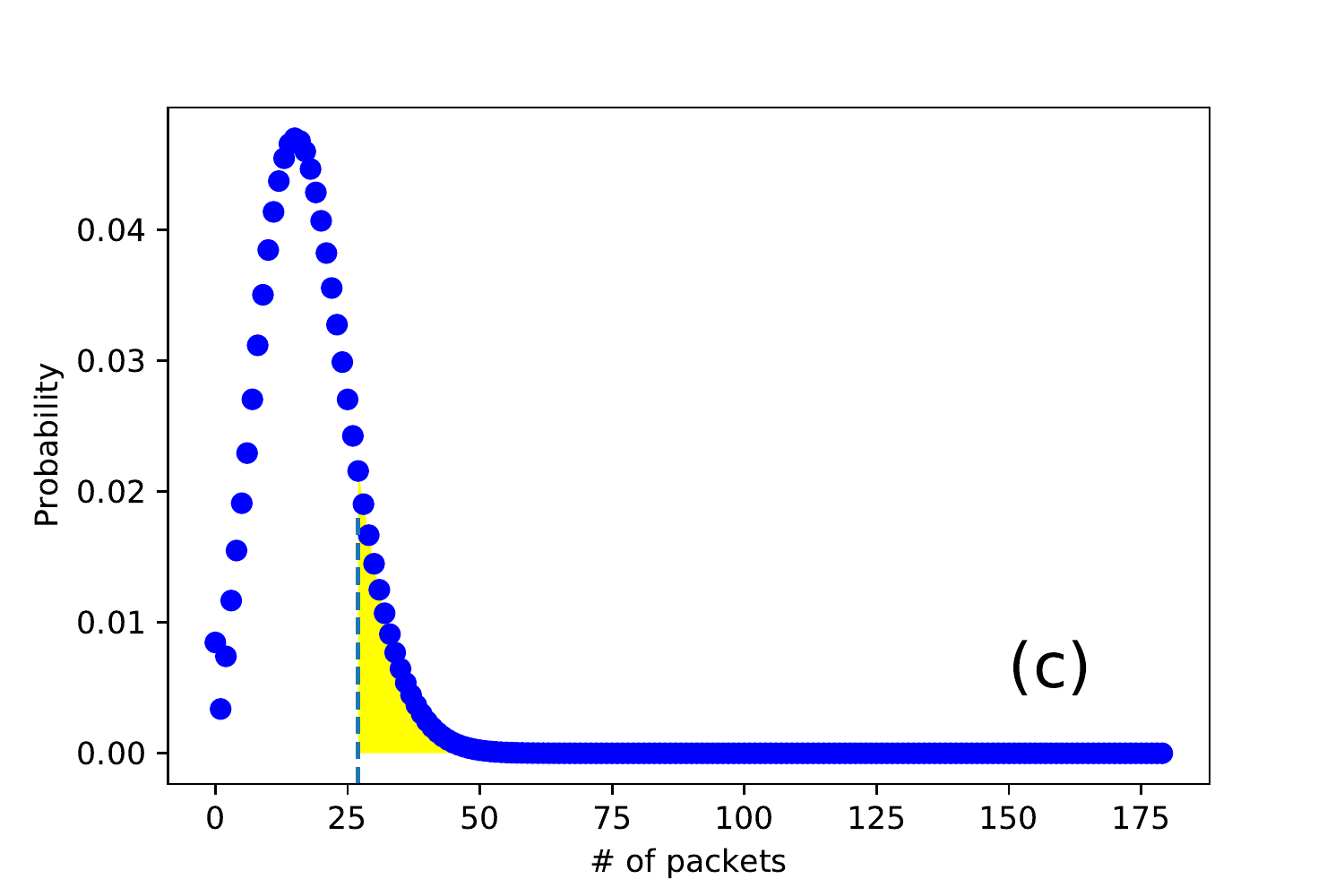}
    \hspace{30px}
    \includegraphics[width = 0.5 \linewidth]{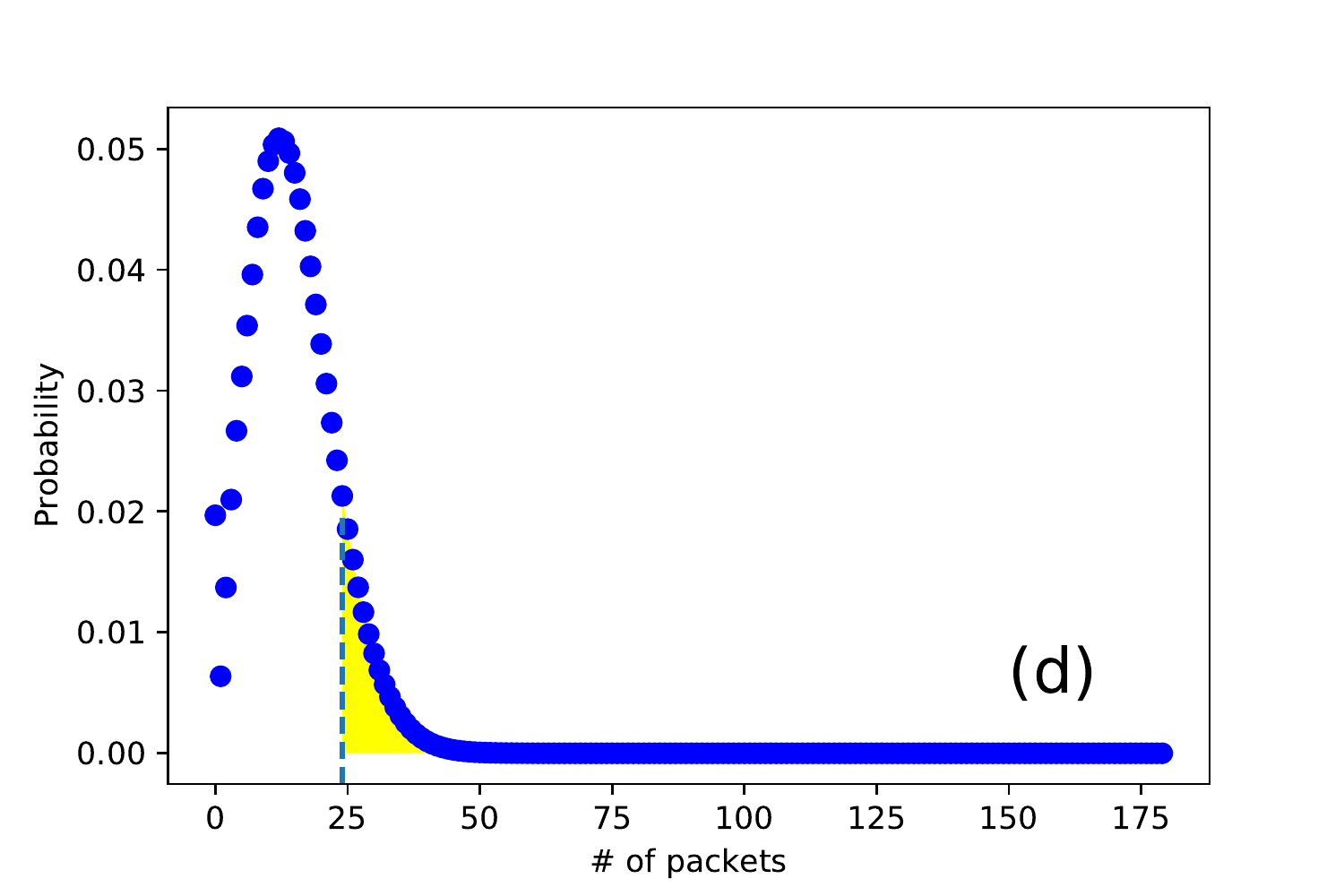}%
    \caption{The evaluation of optimal link capacity for the most central and second most central edge of a B\'{a}rabasi Albert Network of size $30$ and minimum degree $4$. The dotted line in each subfigure represents the value of $l_{ij}$ for the concerned link and the region in yellow to the right of $l_{ij}$ represent all the $x$ values of the PMF the sum of whose corresponding $y$ values is $1-c$ = $1 - 0.85$ = $0.15$ The PMF $\Pi$ and optimal link capacity $l$ for (a) The most central link at $q$ = $1$ (b) The second most central link at $q$ = $1$, (c) The most central link at $q$ = $0.25$ and (d) The second most central link at $q$ = $0.25$ are shown here}
    \label{Fig2}
\end{figure}

The proposed design of our algorithm is to determine pmf. That is based on the concept that if a random variable is expressed as the sum of a number of other random variables, and it has the distribution as the linear convolution of the random variables that make up the sum. In our case, the number of packets flowing through a link $ij$ in a time window is the sum of the number of packets flowing through it due to the involvement of all the pairs of nodes in the network that send packets through $ij$. 
\\In order to evaluate desired pmfs $\Pi_{ij}$, we discuss step by step how we computationally evaluate $\Pi_{ij}(k)$ for $k = 0, 1, 2, ...M-1, where, $ M is an integer.

\begin{enumerate}
\item Evaluate the set $A_{ij} = \{ (x,y) \mid x \in V, y \in V, x \neq y, f_{ij}^{xy} \neq 0\}$
\item  For all $(x,y) \in A_{ij}$, evaluate  $\Omega_{ij}^{xy}(\lambda_{xy},q_{xy},f_{ij}^{xy},\sigma)$ for $\sigma$ being an integer lying in the range $[0,Q-1]$ such that:\\\\ $\Omega_{ij}^{xy}(\lambda_{xy},q_{xy},f_{ij}^{xy},Q-1) << \epsilon \Omega_{ij}^{xy}(\lambda_{xy},q_{xy},f_{ij}^{xy},\lambda_{xy})\quad 0 < \epsilon << 1$ \\\\
and store the values derived from equation (2) in the vector \textbf{$\Omega_{ij}^{xy}$}\\ 
\item Perform the convolution of all the vectors \textbf{$\Omega_{ij}^{xy}$} given in equation (2) for all $(x,y) \in A_{ij}$. The resultant finite seized vector of this convolution represents the distribution of $\Pi_{ij}$ with a small magnitude of error ($\epsilon$). 
\end{enumerate} 
The smaller the value of $\epsilon$ is taken, the more accurate will be the result as the number of iterations will be more. We have taken $\epsilon$ = $0.001$ throughout the paper. The small magnitude of error occurs because we convolve finite length vectors (of size Q) $\Omega_{ij}^{xy}$ and hence get back a finite length vector (of size $M = Q \mid A_{ij}\mid$, $\mid.\mid$ representing the cardinality of a set) $\Pi_{ij}$. Since $M$ would ideally be $\infty$, convolutions of finite size vectors always introduces some error. The degree of error naturally decreases with the increase in the value of $Q$. We discuss the magnitude of this error in an example that we  describe below.
\\In Figure 1, we show the evaluated pmf describing the number of packets between each source-destination pair flowing through three links in a Bar\'{a}basi Albert Graph of size 30 nodes \cite{albert2002statistical}. Figure (1) shows the links having the top three values of edge betweenness centrality in the network i.e., these links which are common to the most number of shortest paths across the network. For the sake of simplicity, we assume that $\lambda_{xy} = \lambda$ and $q_{xy} = q$ for all $x,y \in V$. In Figure 1, the pmf of the aforementioned links are shown for values of $q = 1, 0.75, 0.5, 0.25$  and $\lambda = 4$. We take the value of $Q$ to be $60$. By summing up the values of the resultant vectors $\Pi$, obtained by steps mentioned earlier, and comparing the sum to $1$, we find that error is less than $1$ percent (around 0.01 percent) for each case.\\
Now, having developed the steps for computationally evaluating the pmf for any link in the network, in a form of a finite length vector, we propose a statistical approach in the next section to evaluate the capacity for each link in the network provided some statistical requirements (\textit{local performance criteria}) about their performance is specified.

\section{Evaluation of Optimal Link Capacities for Network}
In this section, we discuss a simple procedure of evaluating the optimal capacity of any link in the network from its pmf. Assume that a statistical condition is provided for the link for which we want to evaluate the optimal capacity. This statistical condition for link $ij$ is represented by $c_{ij}$ and is a real number between 0 and 1. The statistical condition $c_{ij}$ is defined as the probability with which the link $ij$ remains congestion free in a time frame. The goal is to evaluate an optimal link capacity that satisfies this statistical condition. In this paper, the term \textit{local performance criteria} is interchangeably used with statistical condition. Note that these criteria are referred to as \textit{local} because they are the statistical conditions of a single link and not the entire network. It is possible that some links in the network have different statistical conditions than the others. However, for the sake of simplicity, while evaluating the optimal link capacity we assume that these statistical conditions are homogeneous over all edges in the network topology. So, $c_{ij} = c$ for all ij, is used in the rest of the paper. The statistical condition approach was motivated by the bandwidth provisioning described in \cite{DBLP:journals/corr/abs-1710-00420}, where the link transparency was selected as the QoS criterion. 

We define a link to be congestion-free in a time frame when the number of packets flowing through this link in that time frame is less than or equal to its assigned capacity. Using the knowledge of \textit{local performance criteria}, the capacities of links are evaluated using the pmf $\Pi_{ij}$ for each link $ij$ in the given network topology. Steps of evaluating pmf for any link is discussed in the earlier section. The number, at which the cumulative mass function (cmf) of $\Pi_{ij}$ attains the value $c$, is the least link capacity for which the performance criteria are fulfilled. Hence, this is the optimal capacity for the link $ij$.\\ 
In Figure 2, we demonstrate this method of evaluating the link capacity from the pmf for the most central and second most central link of B\'{a}rabasi-Albert Network of size $30$ and minimum degree $4$. The measure of centrality used for the link is the betweenness centrality of the link in the network. The performance criteria $c$ is set at 0.85. This implies that we expect all links in the network to be congestion free with probability $0.85$ in a time-frame. So for evaluating optimal link capacity for all the edges we evaluate the cumulative mass functions (cmf) for all links from their pmfs, and select the numbers for which the cmfs attains $c$ as the optimal capacities of the links.\\
The typical routing algorithm (e.g., OSPF routing protocol for communication networks) considered for the traversal of packets in the shortest path is employed here, it is evident from Figure 2 that the higher edge-betweenness centrality link has the higher (in fact, the highest in the network) optimal capacity after the allocation having done. Also, as $q$ decreases, so does the value of optimal capacity $l_{ij}$ due to change in terms of $\Pi_{ij}$.\\

From Figure 2, it is observed that the qualitative shape of the pmf varies substantially with the changes in the values of $q$. This also causes a change the value of the optimal link capacity. The optimal capacity falls with the changes of values of $q$. This observation also makes sense intuitively.\\
It can also be observed that with the increase of q from $0$ to $1$, the value of the standard deviation of the pmfs for each link falls, making the curve narrower around the mean. This behaviour is shown in Figure 4(f) (given in Section 5). Lower $q$ results in  a flatter pmf while a higher $q$ results in a narrow pmf. As a result, the optimal capacity for a link is farther away from the mean packet flow rate for a low value of $q$ when compared to a high value of $q$. This indicates that if there is a strategy that assigns capacities to links closer to the mean rate of packet flow through a link, it is not always appeared to be optimal. This observation provides a strong argument that a mathematically motivated allocation strategy, like the one presented in this paper, is essential to assign capacities to links. If one assigns capacities to links equal to the mean number of packets flowing through them, then for satisfying the \textit{statistical conditions}, one would suffer inconsistent error for low and high values of $q$. The error is higher for $q$ nearer to $0$ than to $1$.

\section{Global Performance of the Optimal Capacity Allocation and Results}
After having allocated the link capacities by the steps that have been discussed in Sections 3 and 4, it is necessary to quantify a measure of success, globally, so that we can estimate the efficiency of our allocation strategy from the macroscopic perspective of the entire network topology. We have observed through simulations (using the model described in section 2) tested in this work that all links in the network, on an average, remain congestion free for $c$ fraction in total number of time frames. This result from the simulation is not unexpected as we allocate capacities to links in such a way that the local statistical condition of each link is satisfied. 
\begin{figure}[t!]
\includegraphics[width = 0.5 \linewidth]{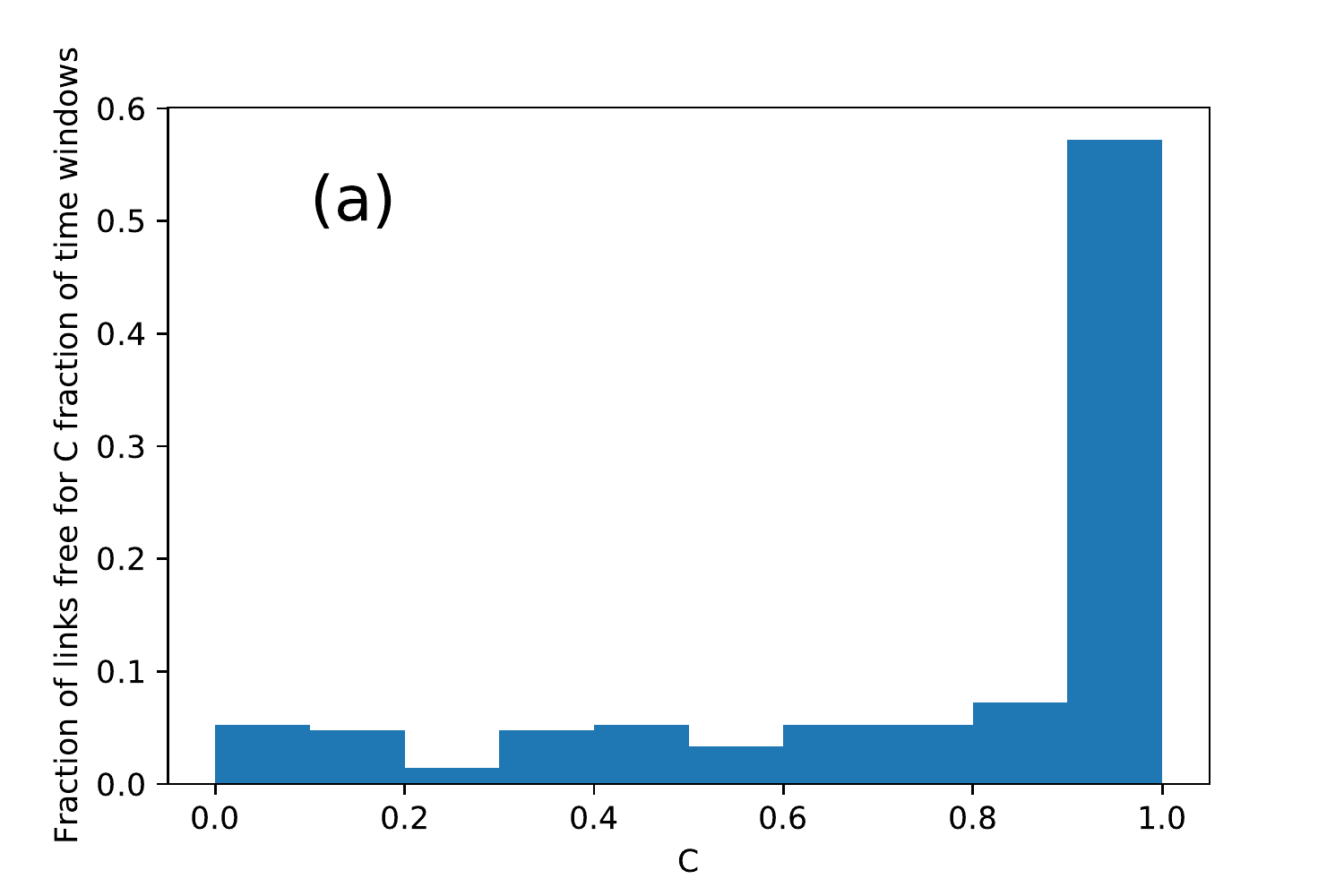} 
    \hspace{30px}
    \includegraphics[width = 0.5 \linewidth]{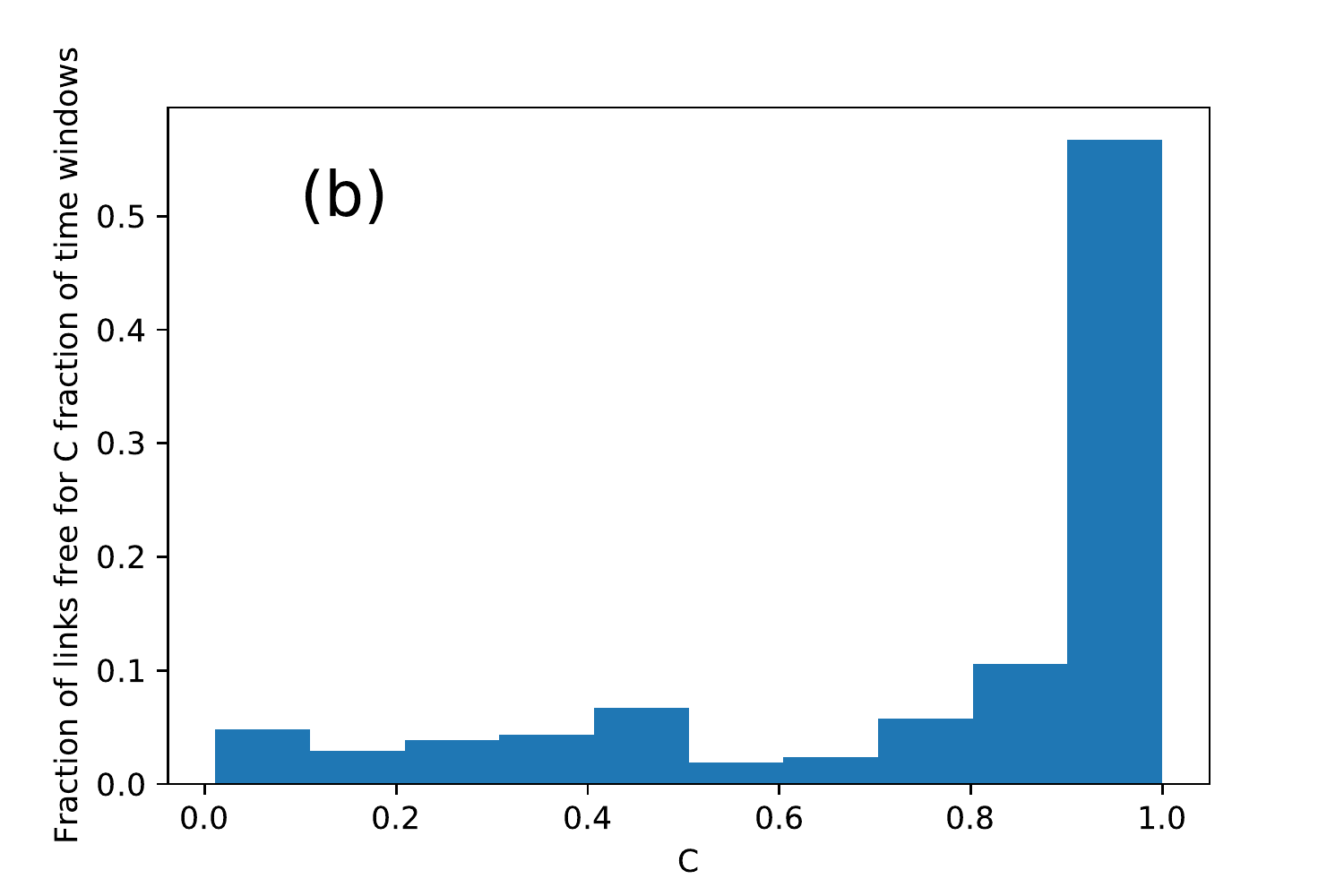}
    \hspace{30px}
    \includegraphics[width = 0.5 \linewidth]{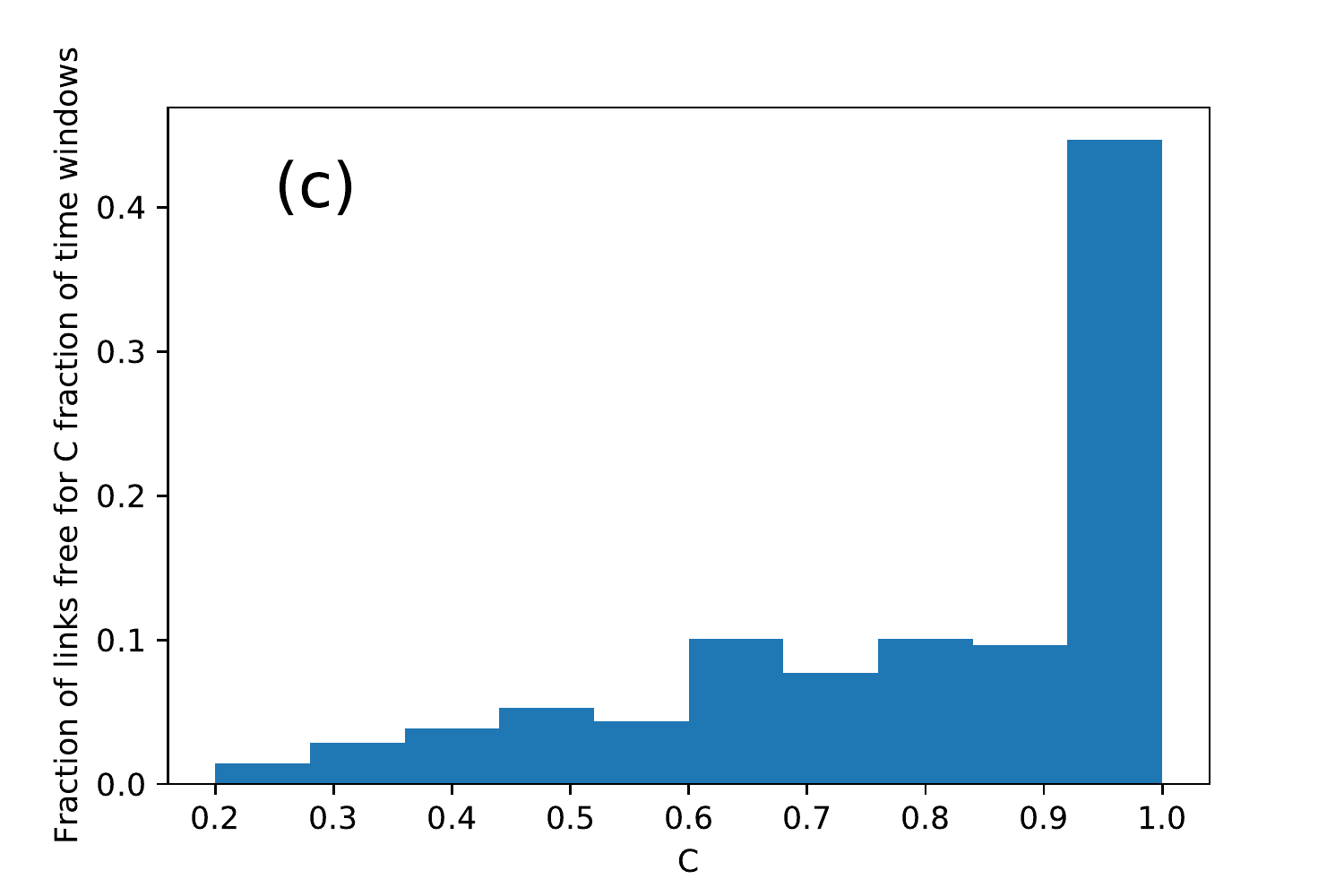}
    \hspace{30px}
    \includegraphics[width = 0.5 \linewidth]{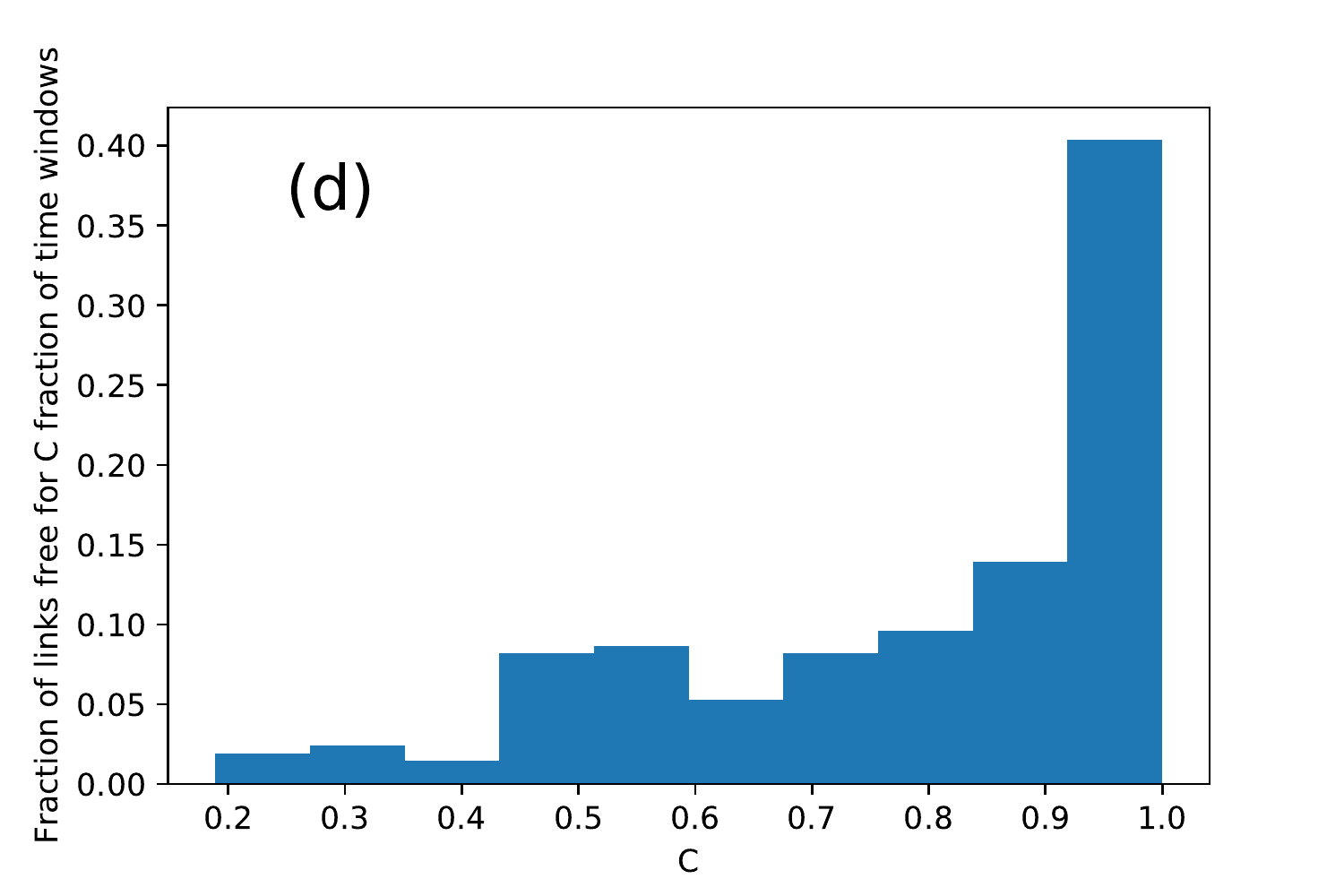}
    \hspace{30px}
    \includegraphics[width = 0.5 \linewidth]{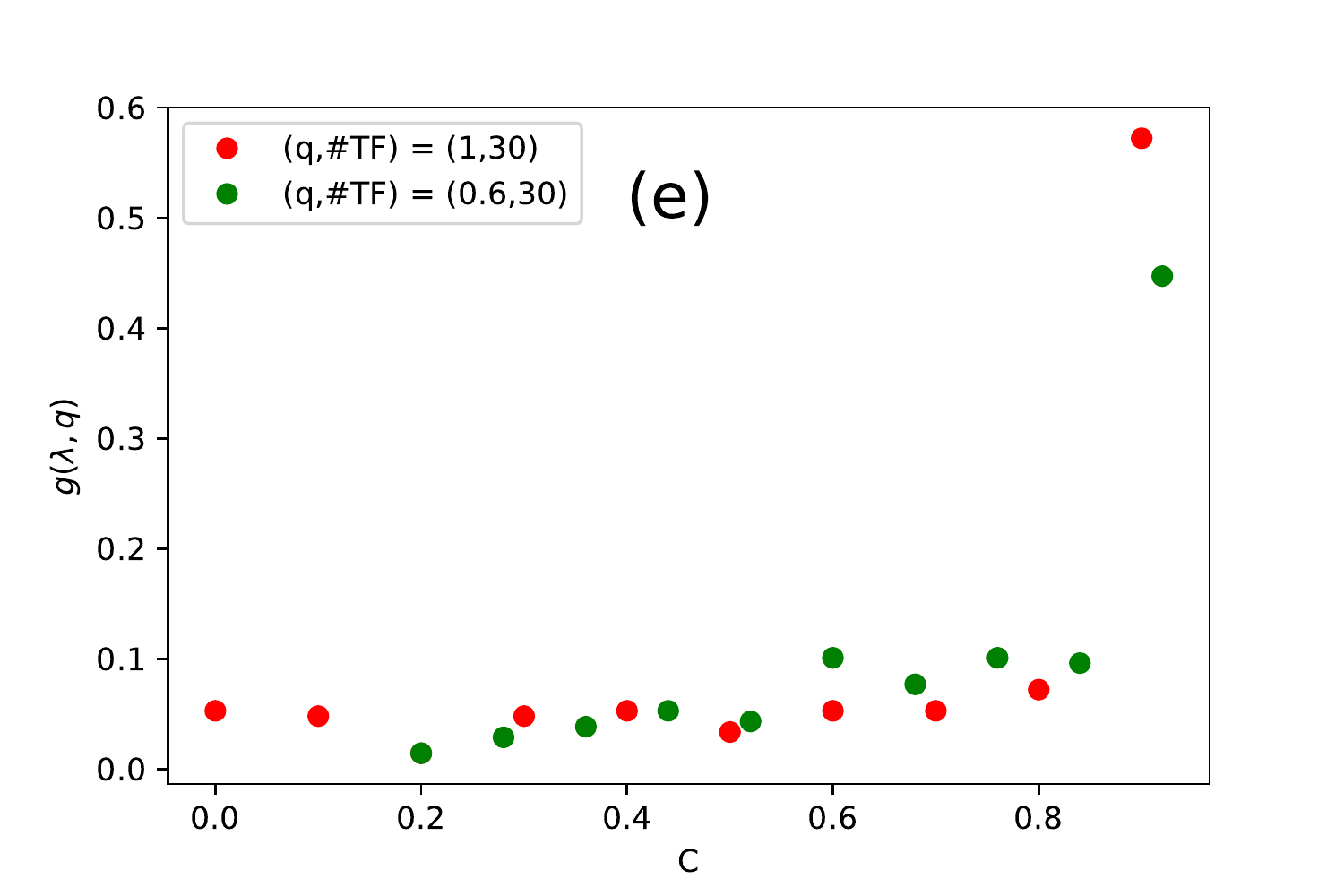}
    \hspace{30px}
    \includegraphics[width = 0.5 \linewidth]{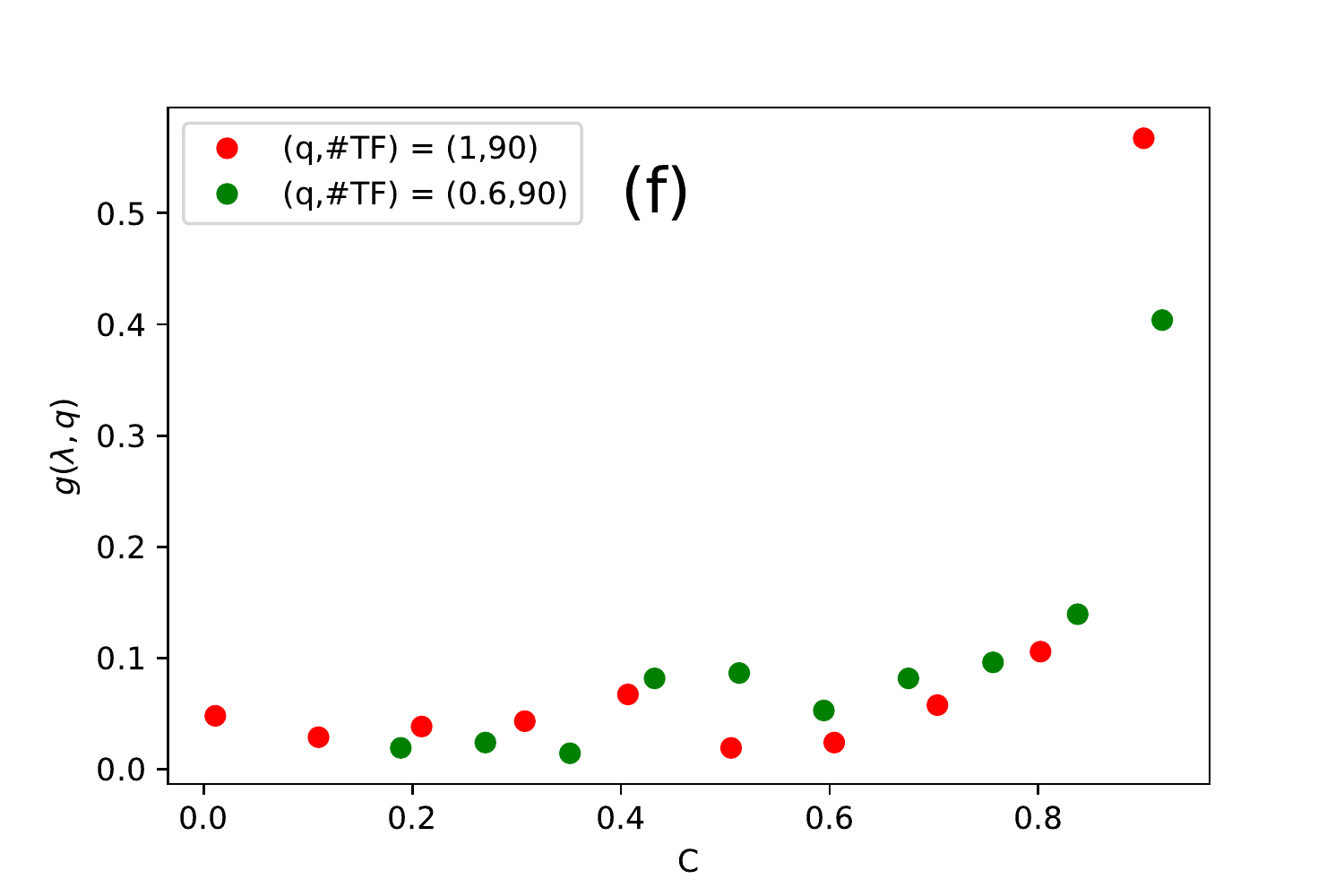}
    \caption{Normalized histogram of the number links congestion - free for C fraction of time frames plotted versus C.The simulation is run for $\lambda$ = 4 and (a) $q = 1$ and 30 time frames (b) $q$ = 1 and 90 time frames (c) $q$ = 0.6 and 30 time frames (d) $q = 0.6$ and 90 time  (e) $g(\lambda = 4,q = 1, 0.6)$ vs $C$ plot for 30 time frames (f) $g(\lambda = 4,q = 1, 0.6)$ vs $C$ plot for 90 time frames. All diagrams with G: Barab\'{a}si Albert Graph of size 30 and minimum degree = 4. }
\label{Fig3}
\end{figure}

The performance, we introduce in this section, is a \textit{global} measure in comparison to observing each link separately, and determining the fraction of observed time the link stays congestion-free. We define the global performance of capacity allocation model as the expected fraction of links in the given network remain congestion-free for $C$  or more fraction of the observed time, for a given set of operating parameters of the underlying process. Note that the fraction $C$, we use for global measure, is different and independent from the local performance criteria $c$.   
\\ 
By operating parameters we mean the parameters which define the underlying traffic flow model. These are $\lambda_{ij}$, $q_{ij}$ $\forall i,j \in V$, $i \neq j$ and $c$. 
From the design point of view, it is preferable to maximize this global measure. It is intuitive that if $C$ is kept constant (say 80 percent of total number of observed time frames), and the value of the local performance criteria $c$ is increased, the value of this global measure would increase as the probability that any randomly selected link in the network is congestion free will increase simultaneously. In this section the global measure is investigated from the following perspectives: (a) For a given network topology, keeping $\lambda$, $q$ and $c$ constant we want to observe how the global measure changes with $C$ and (b) keeping $c$ and $C$ constant and assuming standard probability distributions for $\lambda$ and $q$, we want to observe how the expected value of the global measure varies with the changes of the properties of graph (network topology)like the mean node degree, the mean node centrality, the standard deviation of node degree, the standard deviation of node centrality and the number of edges. 
\\
The global measure or the global performance measure is represented by $g$. This measure $g$ is clearly a function of $\lambda$, $q$, $G$, $c$, and $C$. When the global measure is investigated from the second perspective, $c$ and $C$ are kept constant at 0.85 and 0.8 respectively throughout the simulations. The measure $g$ effectively becomes a function of $\lambda$ and $q$, for a given network topology and it can be represented as $g(\lambda,q)$.
\\
The Figures 3(a)-(d) are shown the normalized histogram of the fraction of links that remain congestion-free for $C$ fraction of observed time frames. We consider that $\lambda$ is constant at 4 and $c$ is at 0.85. Figures 3(a) and 3(b) are exhibited for $q = 1$, while Figures 3(c) and Figure 3(d) are presented for $q = 0.6$. The pairs of Figures 3(a), 3(b) and Figures 3(c), 3(d) vary with the number of time frames on which the topology is observed for. The histogram interval for all the figures 3(a)-(d) is $0.1$. For evaluating $g$ for a particular value of $C$ from any one of these histograms (note that $\lambda$, $q$, $c$ and total number of time frames are constants for each histogram), we must add up the y-values of the histogram bars from $C$ to 1. This summation operation is represented in Figures 3(e) and 3(f). The Figures 3(e) and 3(f) are demonstrated $g(\lambda, q)$ versus $C$ for a set of parameters like $\lambda$ = 4, $q$ = 1.0 and 0.6 and $c$ = 0.85. 
\begin{figure}[h!]
	\includegraphics[width = 0.5 \linewidth]{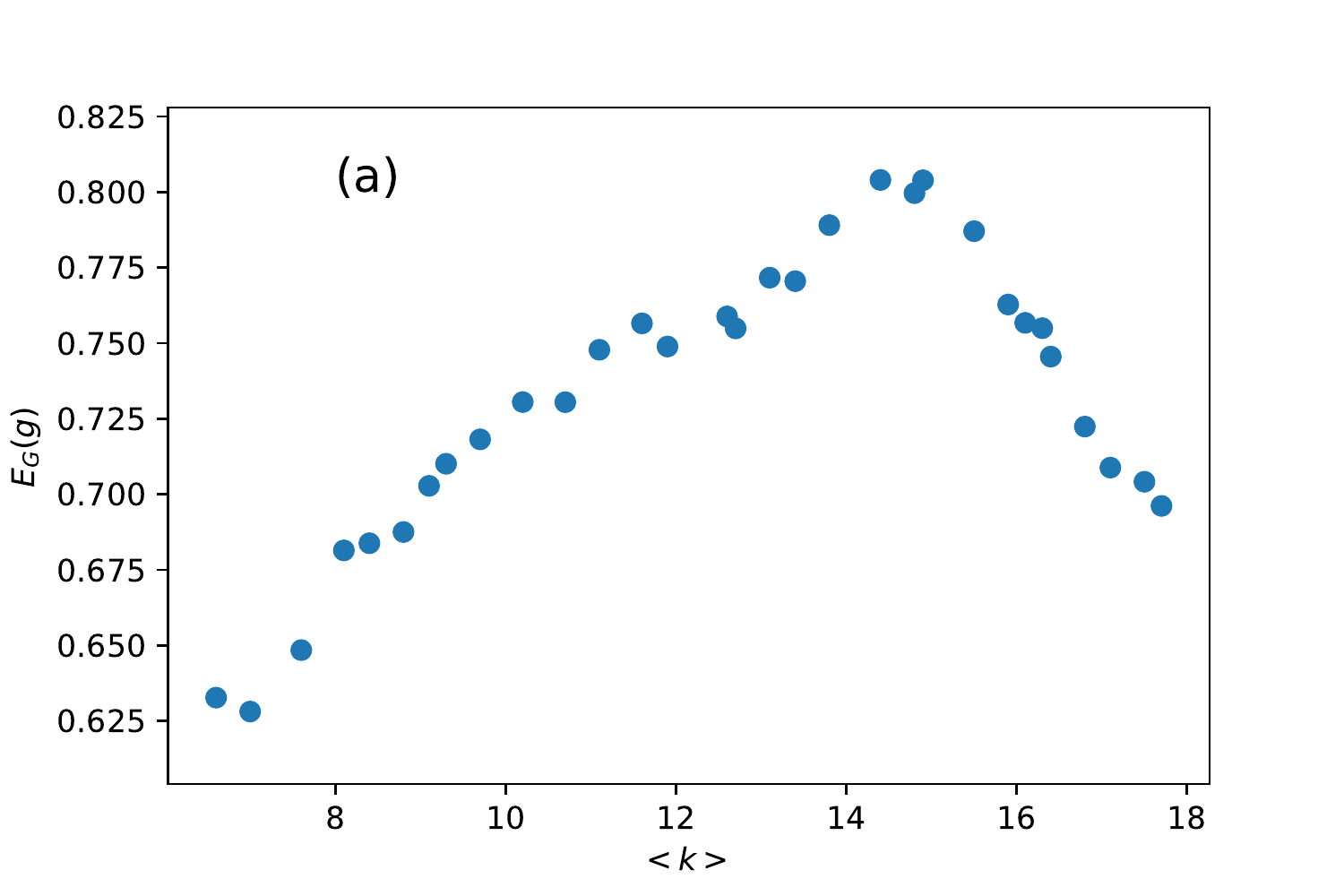} 
    \hspace{30px}
    \includegraphics[width = 0.5 \linewidth]{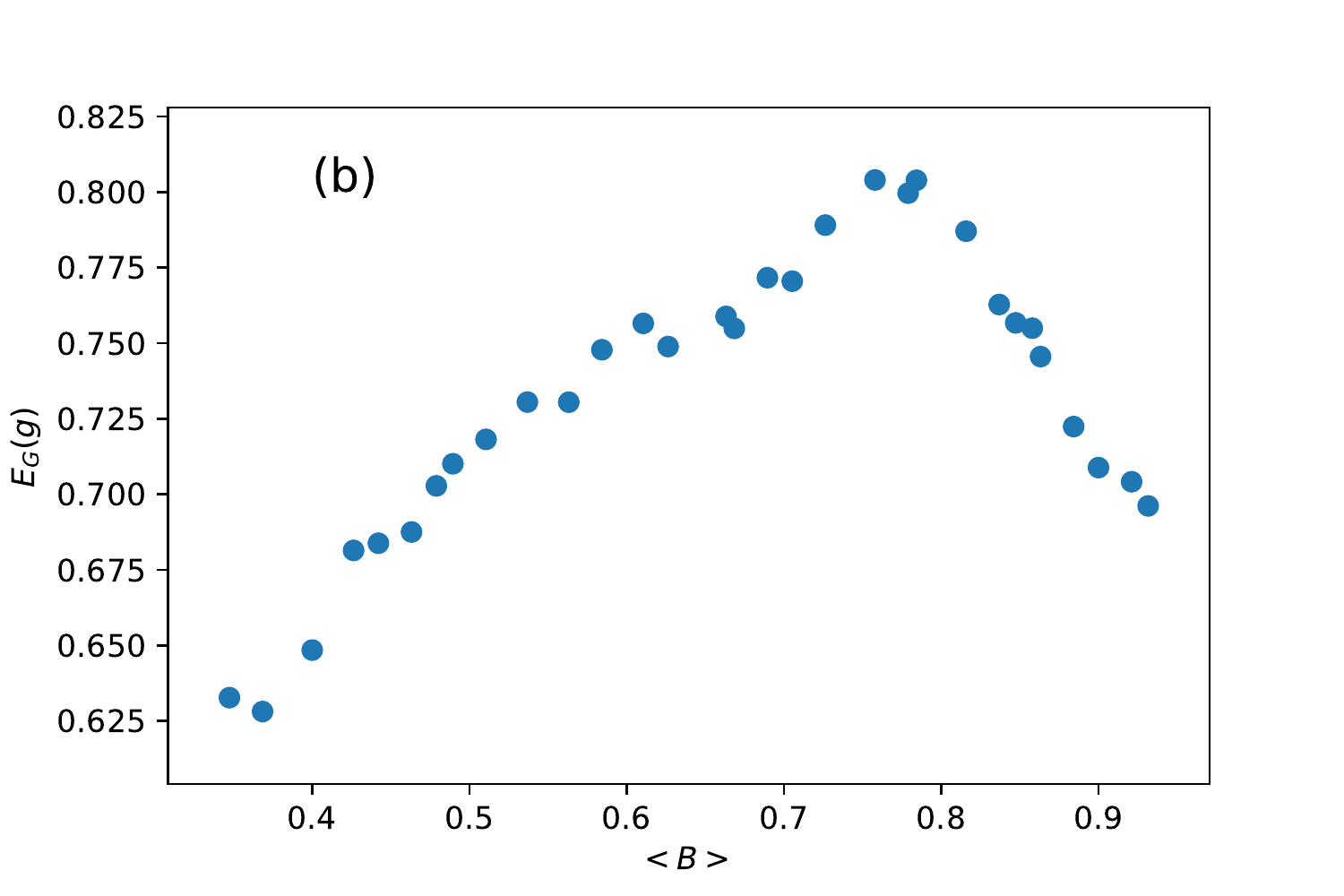}
    \hspace{30px}
    \includegraphics[width = 0.5 \linewidth]{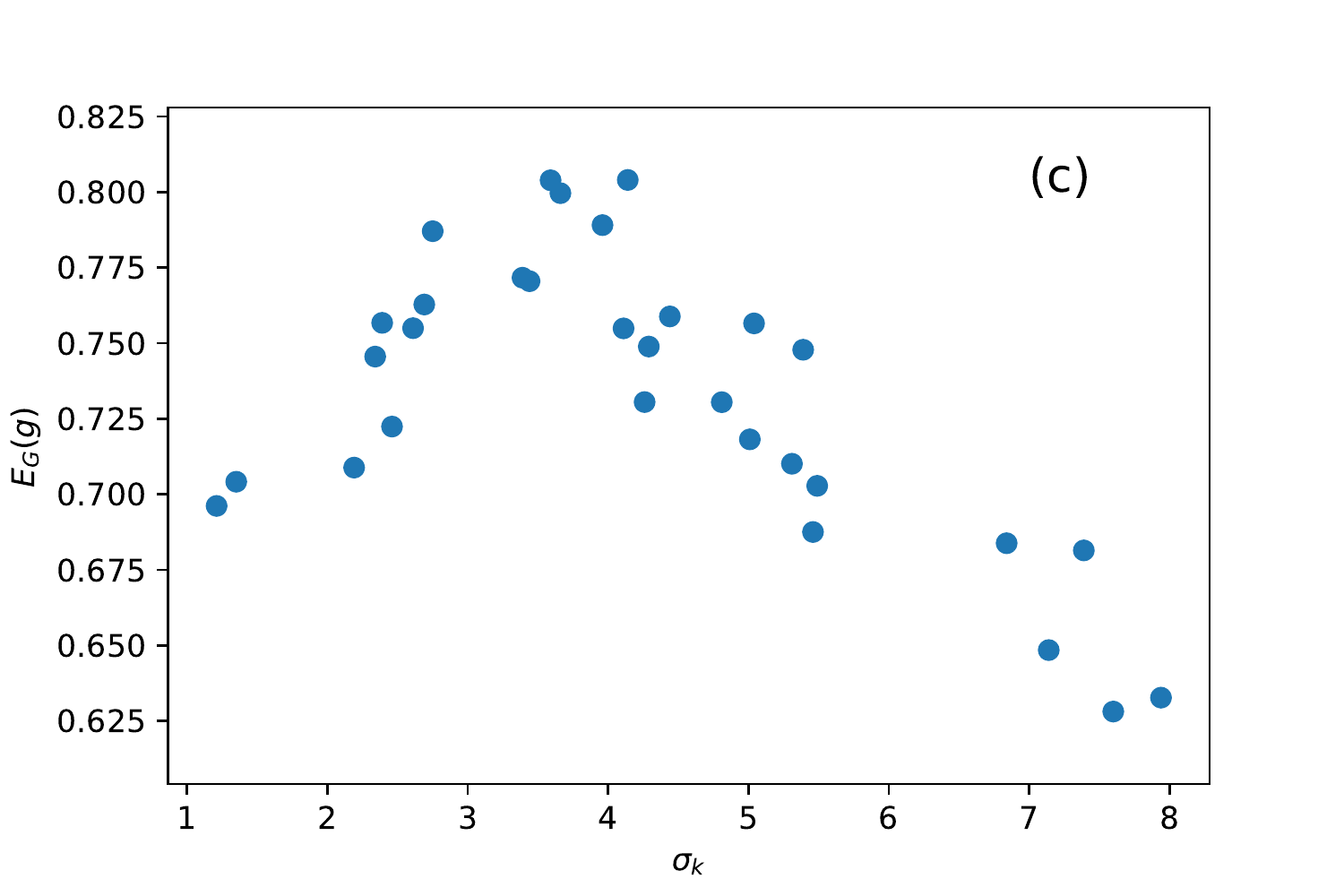}
    \hspace{30px}
    \includegraphics[width = 0.5 \linewidth]{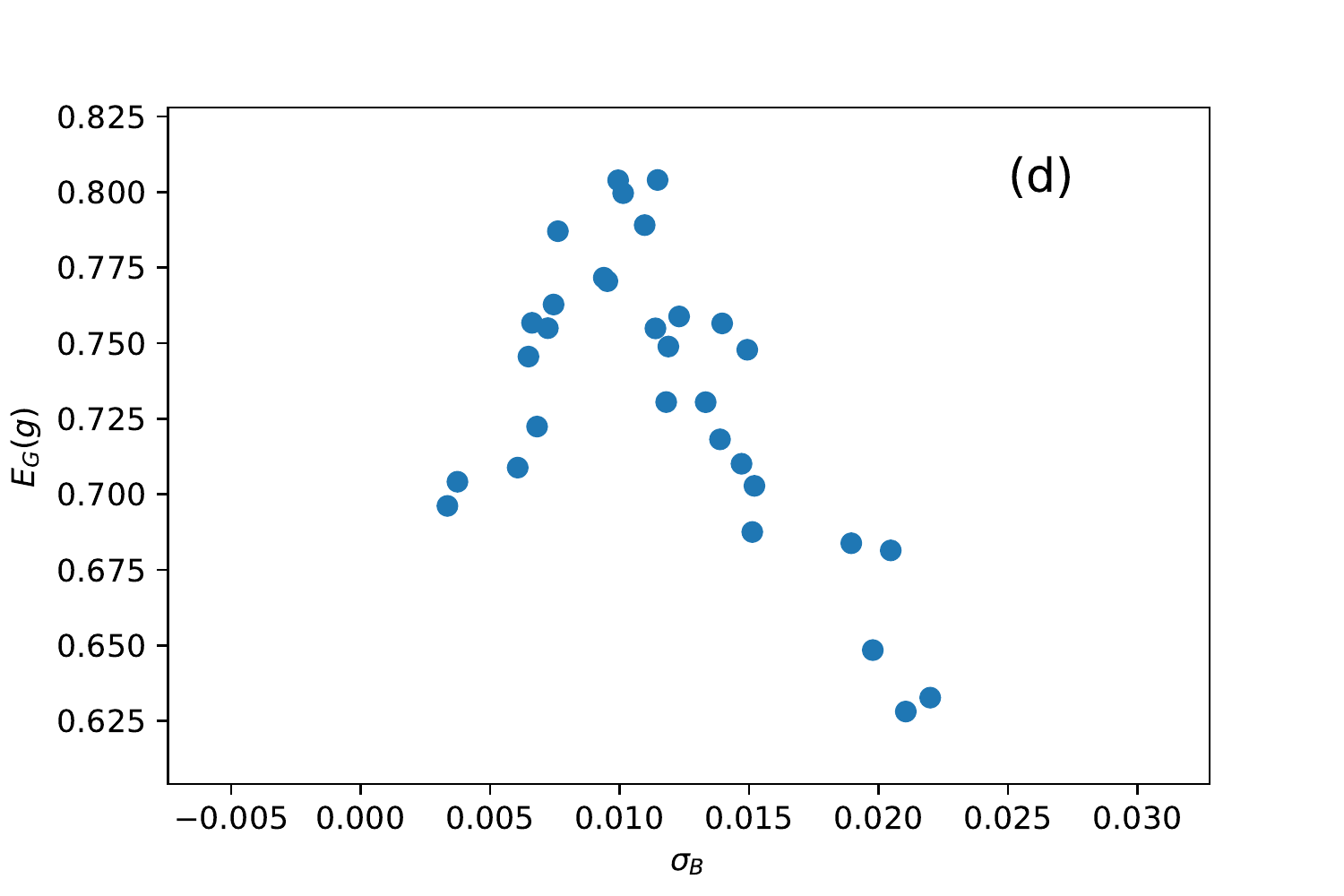}
    \hspace{30px}
    \includegraphics[width = 0.5 \linewidth]{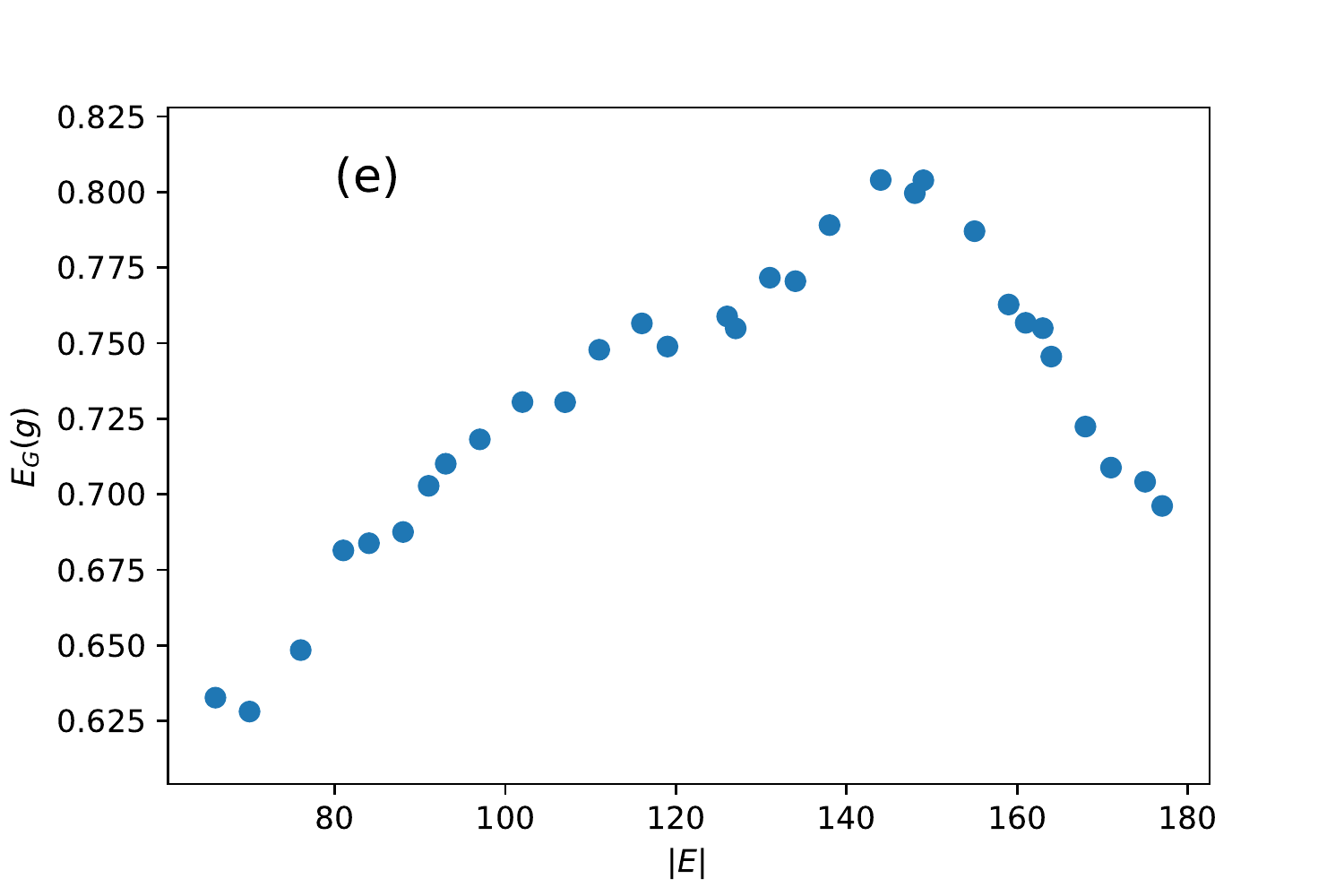}
    \hspace{30px}
    \includegraphics[width = 0.5 \linewidth]{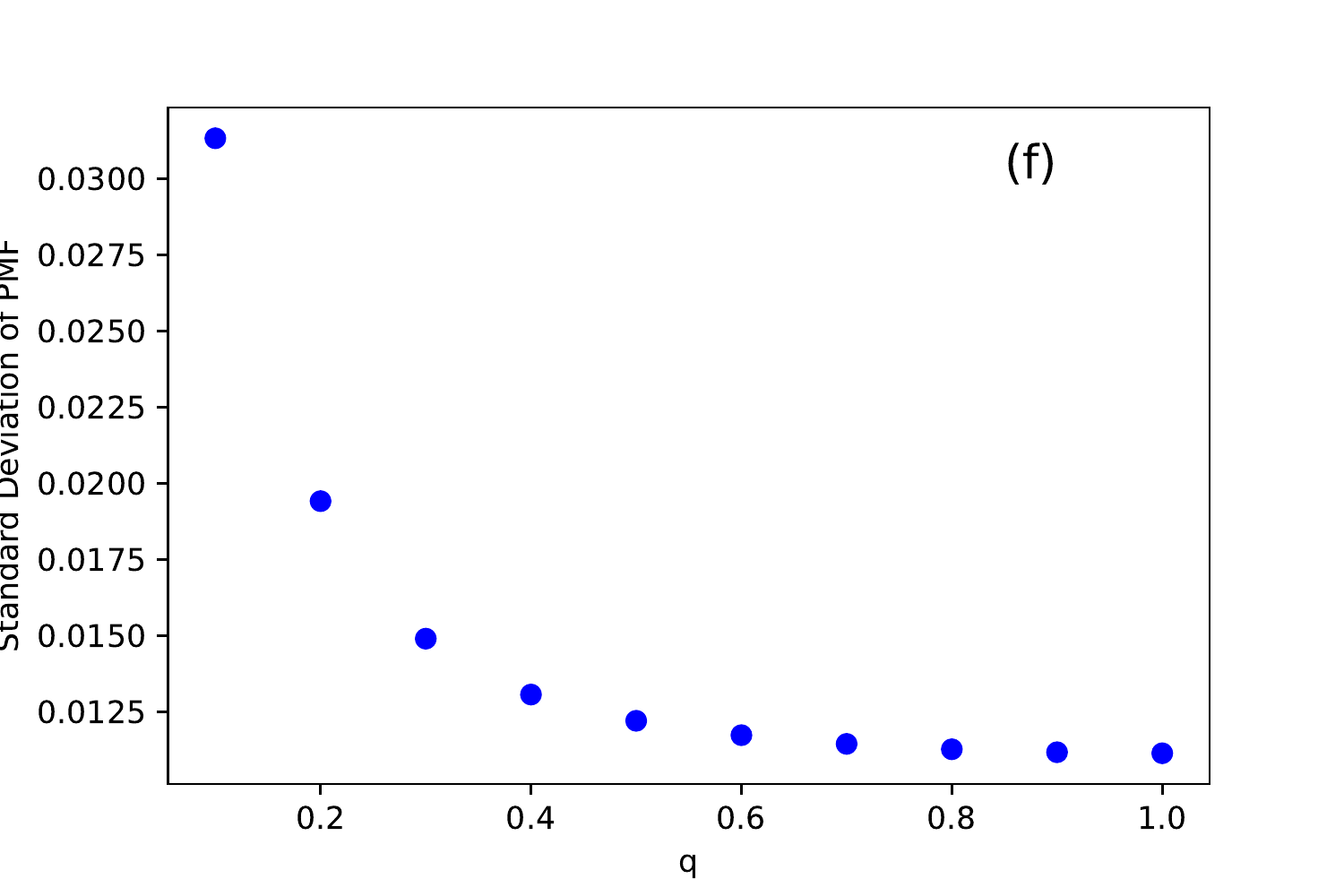}
    \caption{Diagrams of $E_G(g)$ versus (a) Mean Degree $\langle k \rangle$, (b) Mean Centrality $\langle B \rangle$ (c) Degree Standard Deviation $\sigma_k$ (d) Centrality Standard Deviation $\sigma_B$ and (e) Number of Edges in the network topology (f) Standard Deviation of the PMF for maximum edge-betweenness centrality link versus $q$ for $\lambda$ = 4}
\label{Fig4}
\end{figure}
\\
In order to investigate the global measure from the second perspective expected global measure is defined and represented as $E_G(g)$. This is the expected value of global measure of our proposed allocation strategy over all values of $\lambda$ and $q$ for a given network topology. In order to evaluate $E_G(g)$, there must be a priori assumptions of the probability distribution of underlying system parameters namely $\lambda$ and $q$. We do not worry assuming distribution for $c$ or $C$ as these are the performance and design criteria.
If the probability distributions for $\lambda$ and $q$ are represented as $p_{\lambda}$ and $p_q$ then mathematically, $E_G(g)$ can be written as:
\begin{equation}
E_G(g) = \sum_{\lambda = 0}^{\infty}\int_{0}^{1}g(\lambda,q)p_\lambda p_q dq
\end{equation}
For results shown in Figure 4, we assume $p_\lambda$ to be the Poisson Distribution with mean $4$ and $p_q$ to be uniform random density distribution.
For generating the results in Figure 4, a fully connected network of 20 nodes is taken and edges are removed randomly at every step until the network topology was broken down into disjoint sub-graphs. For each network topology generated after the execution of every step, the expected value of global measure $E_G(g)$, is calculated numerically using equation (3) and hence simulation results. The expected values of the measure for different network topologies was drawn with the mean-degrees of these networks. From Figure 4, it is observed that there is a \textit{winner} topology among all of possible connected topologies generated for 20 nodes (for which the value of $E_G(g)$ is maximum at $0.804$). From Figure 4(e) and other similar results from simulations run by changing the mean of $p_\lambda$, we conclude that while removing links randomly from a fully connected network topology of any size, either the \textit{winner} network topology is the graph that is all-connected or it is the graph that arises after a \textit{fraction} of edges is removed randomly from the original all connected network topologies. The \textit{fraction} of edges that need to be removed for arriving at the \textit{winner} topology is entirely depended on the distributions $p_\lambda$, $p_q$ and on the values of local and global performance criteria $c$ and $C$. For cases, when the \textit{winner} network topology is not the initial all-connected graph, expected value of global measure increases as edges are removed until the network reaches a critical point. Beyond that, removal of edges leads to reduction in the expected value of the global measure. 
\section{Discussions on Future Works}
In later works, we would explore in detail whether there exists a finite set of properties of a network topology that determines whether it will be turned out as the \textit{winner} topology among all possible connected topologies with the same number of nodes. With this there can be a clearer understanding of what sort of connectedness leads to maximum efficiency of the proposed allocation strategy and whether there exists an algorithm to generate that \textit{winner} network topology provided that we are given a set of nodes to propose our design with. 
\\The long term goal of our later works will be to build an optimal link capacity allocation algorithm for competent networks supporting any dynamic and greedy routing protocols using the same principles that were used in this work. We aim to proceed by mathematically estimating pmf of the number of packets flowing  through a link in the network for any dynamic and greedy routing protocols. 


\bibliography{ms.bbl}

\end{document}